\documentclass[twocolumn,floatfix,preprintnumbers,nofootinbib,superscriptaddress]{revtex4}

\usepackage{amsmath}
\usepackage{graphicx}
\usepackage{textcomp}
\usepackage{amsfonts}
\usepackage{amssymb}
\usepackage{bm}
\usepackage[dvips]{color}
\usepackage{multirow}
\usepackage{booktabs}
\usepackage[utf8]{inputenc}

\begin{document}

\title{The axial charges of proton within an extended chiral constituent quark model}

\author{Jin-Bao Wang}
\affiliation{School of Physical Science and Technology, Southwest
University, Chongqing 400715, China}

\author{Gang Li}\email{gli@qfnu.edu.cn}
\affiliation{College of Physics and Engineering, Qufu Normal
University, Qufu 273165, China}

\author{Chun-Sheng An}\email{ancs@swu.edu.cn}
\affiliation{School of Physical Science and Technology, Southwest
University, Chongqing 400715, China}

\author{Ju-Jun Xie}\email{xiejujun@impcas.ac.cn}
\affiliation{Institute of Modern Physics, Chinese Academy of
Sciences, Lanzhou 730000, China} \affiliation{School of Nuclear
Science and Technology, University of Chinese Academy of Sciences,
Beijing 101408, China} \affiliation{School of Physics and
Microelectronics, Zhengzhou University, Zhengzhou, Henan 450001,
China}

\thispagestyle{empty}

\date{\today}


\begin{abstract}

We have performed a study of the isovector, octet and singlet axial
charges of the proton in an extended chiral constituent quark model,
where all the possible $uudq\bar{q}$~($q=u,d,s$) five-quark Fock
components in the proton wave function are taken into account. The
$^3P_0$ quark-antiquark creation mechanism is assumed to account for
the transition coupling between three- and five-quark components in
proton, and the corresponding transition coupling strength is fixed
by fitting the intrinsic sea flavor asymmetry $\bar{d}-\bar{u}$ data
for proton. Accordingly, with all the parameters fixed by empirical
values, the probabilities of the intrinsic five-quark Fock
components in proton wave function should be $\sim30 - 50\%$, which
lead to the numerical results for quark spin $\Delta u$, $\Delta d$
and $\Delta s$, as well the axial charges of proton consistent with
the experimental data and predictions by other theoretical
approaches.

\end{abstract}

\maketitle


\section{Introduction}
\label{intro}

Study of the structure of nucleon and nucleon excitations is one of
the most important topic in hadronic physics. Experimentally, most
of the explicit information about proton structure is from the deep
inelastic scattering measurements. Two renowned measurements
performed by European Muon Collaboration
(EMC)~\cite{Ashman:1987hv,Ashman:1989ig} and New Muon Collaboration
(NMC)~\cite{Arneodo:1994sh}, showed us that, in addition to the up
and down quarks, there are also other contributions to the proton
spin. Furthermore, the anti-up and anti-down quarks are asymmetry
inside the proton. Later, there have been also other experimental
efforts dedicated to the measurements of the proton
structure~\cite{Amaudruz:1991at,Arneodo:1994sh,McGaughey:1992kz,Baldit:1994jk,Ackerstaff:1998sr,Anthony:2000fn,Towell:2001nh}.
Very recently, new experimental results by SeaQuest Collaboration
confirmed precisely that the distributions of anti-up and anti-down
quarks inside proton are considerably different and there are more
anti-down quark than anti-up quark over a wide range of
momenta~\cite{Dove:2021ejl}.

On the other hand, one can also use the spin dependent structure
function $g_1(x)$ of the proton to investigate the contribution of
the spin of the quarks to the proton spin. It also relates the
integral over all $x$ of the difference of $g_1(x)$ for the proton
to the scale-invariant axial charges $g_A$ of the target proton.
Indeed, in Ref.~\cite{Ashman:1989ig,Aidala:2012mv}, it was shown
that the flavor-singlet, isovector, and ${\rm SU(3)}$ octet axial
charges of proton, can be obtained by combining the deep inelastic
scattering (DIS) data with the nucleon and hyperon $\beta$-decay
data, which are: $ g^{(0)}_A = 0.120\pm0.093\pm0.138$, $g^{(3)}_A = 1.254 \pm 0.006$, $ g^{(8)}_A =
0.688\pm0.035$. However, these
values are obviously different with the constituent quark model
predictions which are $g^{(0)}_A=1$, $g^{(3)}_A=5/3$, and $g^{(8)}_A=1$, respectively. In addition, the recent measurements
performed by COMPASS
collaboration~\cite{Alexakhin:2006oza,Alekseev:2010hc} showed that $g^{(0)}_A = 0.33 \pm 0.03$,
$g^{(3)}_A = 1.2670 \pm 0.0035$, and $g^{(8)}_A = 0.58 \pm 0.03 \pm
0.05$. Detailed reviews about the
experimental and theoretical progress on proton spin puzzle and the
intrinsic sea flavor asymmetry in proton, were made in
Refs.~\cite{Kuhn:2008sy,Burkardt:2008jw,Chang:2014jba,Leader:2013jra,Wakamatsu:2014zza,Liu:2015xha,Deur:2018roz}.
The fact that the quark-quark interaction is relatively weak at
large energy and momentum scales, whereas it is clouded by the
increasingly strong interaction at lower energy scales is the reason
why, so far, we have no clear picture for the structure of proton.

In fact, one has to go beyond the original three-quark $qqq$ picture
of proton to explain the experimental measurements mentioned above.
In Ji's sum rule~\cite{Ji:1996ek}, the proton spin should be
decomposed as the inner intrinsic quark spin, orbital angular
momentum, as well as those of the gluon inside proton. All these
decomposed contents have been intensively calculated by Lattice
QCD~\cite{Aoki:1996pi,Hagler:2007xi,QCDSF:2011aa,Yang:2016plb,Alexandrou:2017oeh,Yamanaka:2018uud}.
Phenomenologically, both the proton spin puzzle and the intrinsic
sea flavor asymmetry could be solved within the meson cloud
model~\cite{Myhrer:2007cf,Thomas:2008ga}. It's also shown that the
strangeness spin, strangeness magnetic moment, axial form factors of
nucleon~\cite{Zou:2005xy,An:2005cj,Adamuscin:2007fk,Bijker:2012zza},
and electromagnetic and strong decays of several nucleon
resonances~\cite{An:2008xk,An:2011sb,Li:2005jn,Li:2006nm} could be
described well by considering five-quark Fock components in the
baryons' wave functions. In Ref.~\cite{An:2012kj}, the intrinsic sea
content of proton was investigated by an extension of the
traditional chiral constituent quark model to including the
five-quark components, where the $^{3}P_{0}$ model is adopted to
account for the quark-antiquark pair creation in hadrons, which
could result in the transitions between three- and five-quark
components. Recently, the model of Ref.~\cite{An:2012kj} was
phenomenologically applied to study the quark orbital angular
momentum in proton~\cite{An:2019tld}, where the theoretical
calculations showed that $L_{q} = 0.158\pm0.014$. It was shown that
the study of the intrinsic sea quark content of proton is of great
interest to explore its properties. Consequently, by explicit
considering the contributions of the intrinsic five-quark Fock
components, we employ the extended chiral constituent quark model
(E$\chi$CQM) of Ref.~\cite{An:2012kj} to calculate the isovector,
flavor-octet and -singlet axial charges of nucleon in present work.

The present manuscript is organized as follows. In Sec.~\ref{frame},
we give the framework which includes the extended chiral constituent
quark model and the formalism for the nucleon axial charges in
corresponding model, the explicit numerical results are presented in
Sec.~\ref{num}. Finally, a brief summary is given in
Sec.~\ref{conc}.


\section{Framework}
\label{frame}

In this section, we will briefly introduce the E$\chi$CQM
in Sec.~\ref{ecqm}, and present the formalism for calculations of the proton axial charges
within present model in Sec.~\ref{ac}.

\subsection{E$\chi$CQM}
\label{ecqm}

Following Ref.~\cite{An:2012kj}, within the E$\chi$CQM for the
ground state of octet baryons, the wave function for proton can be
expressed as:
    \begin{equation}
|p\rangle=\frac{1}{\sqrt{\mathcal{N}}}\left(|uud\rangle+\sum_{i}C_i^{q}|uudq\bar{q},i\rangle
\right) \label{wfc},
    \end{equation}
where the first term represents the wave function for the
three-quark $uud$ component of the proton, while the sum over $i$
runs over all the possible five-quark configurations with a
$q\bar{q}$ ($d\bar{d}$, $u\bar{u}$, $s\bar{s}$, ...)~\footnote{In
this work, we do not take $c\bar{c}$ and $b\bar{b}$ into account,
since the probabilities for them are much smaller than other
five-quark components inside the proton in the low energy scale.}
pair which may form higher Fock components in the proton,
$C_{i}^q/\sqrt{\mathcal{N}}$ are just the corresponding probability
amplitudes for the five-quark components.

One should notice that the orbital quantum number
of the inner quark and antiquark of the five-quark components in proton
must be odd number $2n+1$ because of the positive parity, while there is
no obvious limit for the radial quantum number of the five-quark system.

\begin{figure}[htbp]
\begin{center}
\includegraphics[scale=0.5]{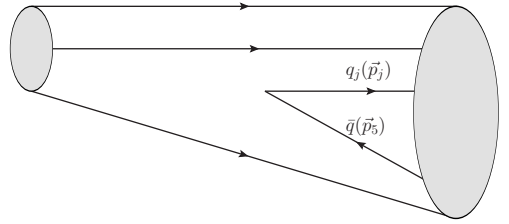}
\end{center}
{\caption{Transition $qqq\rightarrow qqqq\bar{q}$ caused by a
quark-antiquark pair creation in a baryon via the $^{3}P_{0}$
mechanism.} \label{3to5}}
\end{figure}

Once an explicit quark-antiquark pair creation mechanism in the
proton is pinned down, the coefficients $C_i^{q}$ can be calculated
by
\begin{equation}
C_i^q=\frac{\langle uudq\bar{q},i|\hat{T}|uud\rangle}{M_p-E_i},
 \label{coe}
\end{equation}
where $M_p$ is the physical mass of proton, and $E_i$ is the energy
of the $i$th $uudq\bar{q}$ five-quark component. The transition
coupling operator $\hat{T}$ related to the quark-antiquark creation
mechanism, which is taken to be the widely accepted $^3P_0$ coupling
mechanism in the E$\chi$CQM, as shown in Fig.~\ref{3to5}, can be
expressed as
\begin{eqnarray}
\hat{T} &=& -\gamma \sum_{j=1,4}
\mathcal{F}_{j,5}^{00}\mathcal{C}_{j,5}^{00}\mathcal{C}_{OFSC}\sum_{m}\langle
1,m;1,-m|00\rangle \times \nonumber \\
&&
\chi_{j,5}^{1,m}\mathcal{Y}_{j,5}^{1,-m}(\vec{p}_j-\vec{p}_5)b^{\dagger}(\vec{p}_j)d^{\dagger}(\vec{p}_5)
\, , \label{3p0}
\end{eqnarray}
where $\gamma$ is an dimensionless transition coupling constant for
$uud\rightarrow uudq\bar{q}$ , $\mathcal{F}_{j,5}^{00}\text{ and
}\mathcal{C}_{j,5}^{00}$ are the flavor and color singlet of the
created quark-antiquark pair $q_j\bar{q}_5$, $\chi_{j,5}^{1,m}\text{
and }\mathcal{Y}_{j,5}^{1,-m}$ are the total spin $S=1$ and relative
orbital $P-$ states of the created quark-antiquark system, the
operator $\mathcal{C}_{OFSC}$ is to calculate the overlap factor
between the residual three-quark configuration in the five-quark
component and the valence three-quark component, finally,
$b^{\dagger}(\vec{p}_j),d^{\dagger}(\vec{p}_5)$ are the quark and
antiquark creation operators.

Explicit calculations of the matrix elements of the operator
$\mathcal{C}_{OFSC}$, show that the transition coupling between a
$qqqq\bar{q}$ configuration with orbital quantum number $l\ge3$ or
the radial quantum number $n_{r}\ne0$ and the $qqq$ component in
proton vanishes. Consequently, there should be $17$ different
flavor-spin-orbital five-quark configurations which may form Fock
components in the wave function of proton, which are shown in
Table~\ref{con}. In each five-quark configuration, $[v]^F$ denotes
the flavor wave function of four-quark subsystem, and there may be
four different flavor symmetry, i.e. $[v]^F=[31]^{F_1}$,
$[31]^{F_2}$, $[22]^{F}$ and $[211]^{F}$, respectively.

{\squeezetable
\begin{table*}[htbp]
\caption{\footnotesize The orbital-flavor-spin configurations for
five-quark configurations those may exist as higher Fock components
in proton.} \label{con}
\renewcommand
\tabcolsep{0.25cm}
\renewcommand{\arraystretch}{2}
\scriptsize \vspace{0.5cm}
\begin{tabular}{cccccc}
    \toprule[1.2pt]
    $i$  &  $1$  &  $2$  &  $3$  &  $4$  &  $5$  \\
    Config.&$[31]^{\chi}[4]^{FS}[22]^F[22]^S$&$[31]^{\chi}[31]^{FS}[211]^F[22]^S$&$[31]^{\chi}[31]^{FS}[31]^{F_1}[22]^S$&$[31]^{\chi}[31]^{FS}[31]^{F_2}[22]^S$&$[4]^{\chi}[31]^{FS}[211]^F[22]^S$ \\
\hline

    $i$  &  $6$  &  $7$  &  $8$  &  $9$  &  $10$  \\
    Config.&$[4]^{\chi}[31]^{FS}[31]^{F_1}[22]^S$&$[4]^{\chi}[31]^{FS}[31]^{F_2}[22]^S$&$[31]^{\chi}[4]^{FS}[31]^{F_1}[31]^S$&$[31]^{\chi}[4]^{FS}[31]^{F_2}[31]^S$&$[31]^{\chi}[31]^{FS}[211]^F[31]^S$ \\
\hline

    $i$  &  $11$  &  $12$  &  $13$  &  $14$  &  $15$  \\
    Config.&$[31]^{\chi}[31]^{FS}[22]^F[31]^S$&$[31]^{\chi}[31]^{FS}[31]^{F_1}[31]^S$&$[31]^{\chi}[31]^{FS}[31]^{F_2}[31]^S$&$[4]^{\chi}[31]^{FS}[211]^F[31]^S$&$[4]^{\chi}[31]^{FS}[22]^F[31]^S$ \\
\hline

    $i$  &  $16$  &  $17$  &    &   &    \\
    Config.&$[4]^{\chi}[31]^{FS}[31]^{F_1}[31]^S$&$[4]^{\chi}[31]^{FS}[31]^{F_2}[31]^S$&&& \\
 \bottomrule[1.2pt]
\end{tabular}

\end{table*}
}

For the $uudq\bar{q}$ configurations with $[v]^F=[31]^{F_1}$, the
quark-antiquark pair can only be $u\bar{u}$ and $d\bar{d}$. On the
other hand, one can easily obtain that the decomposition between the
flavor wave function for the four-quark subsystem and the antiquark
should be
\begin{eqnarray}
|\frac{1}{2},\frac{1}{2}\rangle_{I}^{[31]^{F_1}} &=&
\sqrt{\frac{2}{3}}|u^3d_{[31]^{F_1}}\rangle\otimes|\bar{u}\rangle +
\nonumber \\
&&
\sqrt{\frac{1}{3}}|u^2d^2_{[31]^{F_1}}\rangle\otimes|\bar{d}\rangle,
\label{31f1}
\end{eqnarray}
where $|\frac{1}{2},\frac{1}{2}\rangle_{I}^{[31]^{F_1}}$ denotes the
isospin state of the proton.

For the $uudq\bar{q}$ configurations with $[v]^F=[31]^{F_2}$, the
quark-antiquark pair can only be $s\bar{s}$, and the corresponding
isospin wave function of the proton is
\begin{equation}
|\frac{1}{2},\frac{1}{2}\rangle_{I}^{[31]^{F_2}}=|u^2ds_{[31]^{F_2}}\rangle\otimes|\bar{s}\rangle
. \label{31f2}
\end{equation}

For the $uudq\bar{q}$ configurations with $[v]^F=[22]^F$,
\begin{eqnarray}
|\frac{1}{2},\frac{1}{2}\rangle_{I}^{[22]^{F}}&=&|u^2d^2,[22]^F\rangle\otimes|\bar{d}\rangle, \\
|\frac{1}{2},\frac{1}{2}\rangle_{I}^{[22]^{F}}&=&|u^2ds,[22]^F\rangle\otimes|\bar{s}\rangle,
\end{eqnarray}
which are the isospin wave functions for the five-quark components
in proton with $d\bar{d}$ and $s\bar{s}$ pairs, respectively.

At last, for the $uudq\bar{q}$ configurations with $[v]^F=[211]^F$,
which limits the quark-antiquark pair to be $s\bar{s}$,
\begin{equation}
|\frac{1}{2},\frac{1}{2}\rangle_{I}^{[211]^{F}}=|u^2ds_{[211]^{F}}\rangle\otimes|\bar{s}\rangle
. \label{211f}
\end{equation}

The total spin for the four-quark system with spin symmetry
$[22]^{S}$ should be $S=0$, therefore, a general wave function for
the five-quark configurations with numbers $i = 1 \cdots 7$ can be
expressed as
\begin{eqnarray}
|p,\uparrow\rangle &=& \sum_{ijkln}\sum_{ab}\sum_{m\bar{s}_z}C^{\frac{1}{2},\uparrow}_{1,m;\frac{1}{2},\bar{s}_z}C^{[1^4]}_{[31]^{\chi FS}_k;[211]^C_{\bar{k}}}C^{[31]^{\chi FS}_{k}}_{[O]^\chi_i;[FS]^{FS}_j} \nonumber \\
&&C^{[FS]^{FS}_{j}}_{[F]^F_l;[22]^{S}_n}C_{a,b}^{[2^{3}]^{C}}|[211]^C_{\bar{k}}(a)\rangle|[11]^{C,\bar{q}}(b)\rangle|\frac{1}{2},\frac{1}{2}\rangle_{I}^{[F]^{F}_{l}} \nonumber \\
&&|1,m\rangle^{[O]^\chi_i}|[22]^S_n\rangle|\bar{\chi},\bar{s}_z\rangle\phi(\{\vec{r}_q\}),
\label{22s}
\end{eqnarray}
where the coefficients $C^{[\cdots]}_{[\cdots][\cdots]}$ represent the CG coefficients of
the $S_{4}$ permutation group, $|[211]^C_{\bar{k}}(a)\rangle$ and $|[11]^{C,\bar{q}}(b)\rangle$
the color wave functions for the four-quark subsystem and the antiquark, combination of which
results in the color singlet $|[2^{3}]^{C}\rangle$ as
\begin{eqnarray}
|[2^{3}]^{C}\rangle&=\sum_{ab}C_{a,b}^{[2^{3}]^{C}}|[211]^C_{\bar{k}}(a)\rangle|[11]^{C,\bar{q}}(b)\,.
\end{eqnarray}
And the state $|1,m\rangle^{[O]^\chi_i}$, which combines to the spin
of the antiquark $|\bar{\chi},\bar{s}_z\rangle$ to form the proton
spin state $|p,\uparrow\rangle$, can be obtained from the
decomposition of the orbital angular momentum of the four-quark
subsystem and the antiquark.

For the configurations with numbers $i = 8 \cdots 17$, the spin
symmetry $[31]^{S}$ of the four-quark subsystem results in the spin
$S=1$, which leads to $J=0$ or $1$ when combining the orbital
angular momentum of the four-quark subsystem to the antiquark. For
the $J = 0$ case, which has been considered explicitly in
Ref.~\cite{An:2012kj}, a general wave function for the 10
configurations can be expressed as
\begin{eqnarray}
|p,\uparrow\rangle &=& \!\! \sum_{ijkln}\sum_{ab}\sum_{ms_z}C^{00}_{1,m;1,s_z}C^{[1^4]}_{[31]^{\chi FS}_k;[211]^C_{\bar{k}}}C^{[31]^{\chi FS}_{k}}_{[O]^\chi_i;[FS]^{FS}_j}\nonumber \\
&& \!\! C^{[FS]^{FS}_{j}}_{[F]^F_l;[31]^{S}_n}C_{a,b}^{[2^{3}]^{C}}|[211]^C_{\bar{k}}(a)\rangle|[11]^{C,\bar{q}}(b)\rangle|\frac{1}{2},\frac{1}{2}\rangle_{I}^{[F]^{F}_{l}} \nonumber \\
&& \!\!
|1,m\rangle^{[O]^\chi_i}|[31]^S_n,s_z\rangle|\bar{\chi},\bar{s}_z\rangle\phi(\{\vec{r}_q\}).
\label{31s0}
\end{eqnarray}
While for the $J = 1$ cases,
\begin{eqnarray}
|p,\uparrow\rangle &=& \!\!
\sum_{ijkln}\sum_{ab}\sum_{J_z\bar{s}_z}\sum_{ms_z}
C^{\frac{1}{2},\frac{1}{2}}_{1,J_z;\frac{1}{2},\bar{s}_z}C^{1,J_z}_{1,m;1,s_z}C^{[1^4]}_{[31]^{\chi FS}_k;[211]^C_{\bar{k}}} \nonumber \\
&&\!\! C^{[31]^{\chi FS}_{k}}_{[O]^\chi_i;[FS]^{FS}_j}C^{[FS]^{FS}_{j}}_{[F]^F_l;[31]^{S}_n}C_{a,b}^{[2^{3}]^{C}}|[211]^C_{\bar{k}}(a)\rangle|[11]^{C,\bar{q}}(b)\rangle \nonumber \\
&& \!\! |\frac{1}{2},\frac{1}{2}\rangle_{I}^{[F]^{F}_{l}}
|1,m\rangle^{[O]^\chi_i}|[31]^S_n,s_z\rangle|\bar{\chi},\bar{s}_z\rangle\phi(\{\vec{r}_q\}).
\label{31s1}
\end{eqnarray}

In present work, we consider both $J=0$ and $J=1$ cases. Hereafter,
we denote these two cases as Set I and Set II, respectively.

Finally, to calculate the energy for a given five-quark
configuration $E_{i}$ in Eq.~(\ref{coe}), we employ the traditional
chiral constituent quark model, in which the quark-quark hyperfine
interaction is~\cite{Glozman:1995fu}
\begin{eqnarray}
H_{hyp} &=& -\sum_{i<j}\vec{\sigma}_i\cdot
\vec{\sigma}_j\Bigg[\sum_{a=1}^3V_\pi (r_{ij})\lambda^a_i\lambda^a_j
\nonumber\\
&& +\sum_{a=4}^7 V_K (r_{ij})\lambda^a_i\lambda^a_j +V_\eta
(r_{ij})\lambda^8_i\lambda^8_j\Bigg]\, ,
\end{eqnarray}
where $\lambda^a_i$ is the Gell-Mann matrix in flavor space which
acts on the $i$-th quark, $V_M (r_{ij})$ is the potential of the
meson \textit{M} exchange interaction between $i$-th quark and
$j$-th quark. Numerical values for all the exchange coupling
strength constants are taken to be the empirical
ones~\cite{Glozman:1995fu}. Then the energies $E_i$ for the 17
five-quark configurations in Table~\ref{con} should be
\begin{equation}
E_{i}=E_{0}+\langle H_{hyp}\rangle + \delta_{q\bar{q}},
\end{equation}
where $E_{0}$ is a degenerated energy for the 17 five-quark
configurations. The parameter $E_0$ is dependent on the constituent
quark masses, the kinetic quark energies, and also the energies of
the quark confinement interactions. Here we take $E_{0}=2127$~MeV,
$\delta_{u\bar{u}} = \delta_{d\bar{d}} = 0$, and $\delta_{s\bar{s}}
= 240$ MeV as used in Ref.~\cite{An:2012kj}.

\subsection{Formalism for the axial charges of proton} \label{ac}

In this section, we present the formalism for the axial charges
$g_{A}^{(0)}$, $g_{A}^{(8)}$ and $g_{A}^{(3)}$ of proton within the
E$\chi$CQM.

It's shown that quark spin contribution $\Delta q$ measured in
polarized DIS is related to the matrix element of the quark axial
vector current operator as
    \begin{align}
    \langle p,s_z|\int\mathrm{d}x\bar{q}\gamma^{\mu}\gamma^5q|p,s_z\rangle&=s^{\mu}\cdot\Delta q\,,
    \end{align}
where $s^{\mu}$ is the proton polarization vector, and $\Delta q$
the difference of quark with spin parallel or antiparallel to the proton spin
\begin{equation}
\Delta q=(q^{\uparrow}+\bar{q}^{\uparrow})-(q^{\downarrow}+\bar{q}^{\downarrow})\,.
\end{equation}
Combinations of different flavor $\Delta f$ with $f=u,d,s$
lead to isovector, flavor octet, and flavor singlet axial charges
of proton,
    \begin{eqnarray}
    g^{(3)}_A &=& \Delta u-\Delta d, \label{g3} \\
    g^{(8)}_A &=& \Delta u+\Delta d-2\Delta s, \label{g8} \\
    g^{(0)}_A &=& \Delta u+\Delta d+\Delta s. \label{g0}
    \end{eqnarray}

In the non-relativistic approximation, $\Delta f$ can be calculated
as:
    \begin{equation}
    \Delta f=\langle ps_z|\sum_{j}\,\hat{\sigma}^z_j\delta_{jf}|ps_z\rangle,
    \end{equation}
where $|ps_z\rangle$ is the wave function Eq.~(\ref{wfc}) of proton,
$\hat{\sigma}^z_j$ is Pauli operator acting on $j$-th quark, and
$\delta_{jf}$ is a flavor-dependent operator. $\delta_{jf}=1$ when
the flavor of $j$-th quark is same with $f$, otherwise,
$\delta_{jf}=0$. Consequently, one can get
 \begin{eqnarray}
 && \!\!\!\!\! \Delta f = \frac{1}{\mathcal{N}}\langle uud,s_z|\sum_{j=1,3}\,\hat{\sigma}^z_j\delta_{jf}|uud,s_z\rangle +  \nonumber \\
 && \!\!\!\!\! \sum_{i}\frac{(C_{i}^{q})^{2}}{\mathcal{N}}\langle
             uudq\bar{q},i,s_z|\sum_{j=1,5}\,\hat{\sigma}^z_j\delta_{jf}|uudq\bar{q},i,s_z\rangle
             ,
             \label{mel}
 \end{eqnarray}
where the non-diagonal terms are assumed to be negligible.

For simplicity, we denote the matrix elements for an explicit given
five-quark component in the second term in Eq~(\ref{mel}) as
    \begin{equation}
    \Delta f_i^q=\langle uudq\bar{q},i,s_z|\sum_{j=1}^5\,\hat{\sigma}^z_j\delta_{jf}|uudq\bar{q},i,s_z\rangle.
    \end{equation}

Finally, one can easily obtain the explicit expression of the matrix
elements results for $\Delta u$, $\Delta d$ and $\Delta s$ of proton
as following
    \begin{equation}
    \Delta u=\frac{1}{\mathcal{N}}\frac{4}{3}+\sum_{i,q}\frac{(C_{i}^{q})^{2}}{\mathcal{N}}\Delta u_i^q, \label{Deltau}
    \end{equation}
    \begin{equation}
    \Delta d=\frac{1}{\mathcal{N}}\left(-\frac{1}{3}\right)+\sum_{i,q}\frac{(C_{i}^{q})^{2}}{\mathcal{N}}\Delta d_i^q, \label{Deltad}
    \end{equation}
    \begin{equation}
    \Delta s=\sum_{i,q}\frac{(C_{i}^{q})^{2}}{\mathcal{N}}\Delta s_i^q. \label{Deltas}
    \end{equation}


\section{The numerical results}
\label{num}

Using the formalism developed in the previous section, here we
present our numerical results for the axial charges of the proton.
Before showing the theoretical results, we firstly discuss the model
parameter $V$, which is the coupling strength for Goldstone boson
exchange. The parameter $V$ can be easily obtained from the matrix
elements of the operator in Eq.~(\ref{3p0}), and it is a common
factor for all the five-quark configurations listed in
Table~\ref{con}, in which the transition coupling constante $\gamma$
is included. As discussed in Sec.~\ref{ecqm}, we have two different
sets in present calculations. For Set I, we just employ the
numerical values for the parameters using in Ref.~\cite{An:2012kj},
where $J=0$ was considered. While for Set II, we keep all the
parameters to be the same values as Set I except for $V$, which is,
now, for the case of $J=1$. For Set II, the value of $V$ is
determined by fitting the sea flavor asymmetry of proton
$\bar{d}-\bar{u}=0.118\pm0.012$~\cite{Towell:2001nh}, and one can
get
\begin{equation}
V = 697 \pm {80} ~~{\rm MeV} \, ,
\end{equation}
where the uncertainty is obtained from the experimental error of the
value of $\bar{d} - \bar{u}$.

Accordingly, explicit calculations result in the numerical results
for the  probabilities of five-quark components $P_{q\bar{q}}$, and
contributions to the quark spin of each five-quark configuration
$\Delta f_{i}^{q}$ shown in Table~\ref{pdelta} for Set I and II,
respectively. Note that we have listed the quark-antiquark pairs in
all the different $uudq\bar{q}$ configurations being light
($u\bar{u}$ or $d\bar{d}$) and strange flavors in two rows denoted
by $l\bar{l}$ and $s\bar{s}$ in the table, respectively.


\begin{table*}[htbp]
\caption{\footnotesize The numerical results of the probabilities $P_{q\bar{q}}$
and the matrix elements of the quark spin $\Delta f_{i}^{q}$ of all the 17 five-quark
configurations in Set I and Set II. Note that we have denoted the five-quark configurations
$uudq\bar{q}$ with light quark-antiquark pairs as $l\bar{l}$, and those with strange
quark-antiquark pairs as $s\bar{s}$.}
\label{pdelta}
\renewcommand
\tabcolsep{0.45cm}
\renewcommand{\arraystretch}{1.8}
\scriptsize
\begin{tabular}{cc|cccc|cccc}
\toprule[1.2pt]

                       &            &    \multicolumn{4}{|c|}{Set I} &  \multicolumn{4}{c}{Set II} \\\hline

Con. $i$               & $q\bar{q}$ & $P_{q\bar{q}} $ & $\Delta u_i^q$ & $\Delta d_i^q$ & $\Delta s_i^q$ & $P_{q\bar{q}}$ & $\Delta u_i^q$ &   $\Delta d_i^q$         & $\Delta s_i^q$ \\
\hline


\multirow{2}{*}{ $1$ } & $l\bar{l}$ & $0.146\pm0.015$ &        $0$     &    $-1/3$      &      $0$       & $0.157\pm0.016$ &   $0$       & $-1/3$                 & $0$            \\

                       & $s\bar{s}$ & $0.010\pm0.001$ &        $0$     &      $0$       &      $-1/3$    & $0.011\pm0.002$ &  $0$        & $0$                    & $-1/3$\\
\hline

\multirow{2}{*}{ $2$ } & $l\bar{l}$ &        $0$      &        $-$     &      $-$       &      $-$       &       $0$       &   $-$       &  $-$                    & $-$ \\

                       & $s\bar{s}$ & $0.004\pm0.001$ &        $0$     &      $0$       &      $-1/3$    & $0.004\pm0.001$  &  $0$       & $0$                    & $-1/3$\\
\hline

\multirow{2}{*}{ $3$ } & $l\bar{l}$ & $0.016\pm0.002$ &       $-2/9$   &     $-1/9$     &        $0$     & $0.018\pm0.002$  &  $-2/9$    & $-1/9$                 & $0$ \\

                       & $s\bar{s}$ &        $0$      &        $-$     &      $-$       &         $-$    &         $0$      &    $-$     &      $-$                    &      $-$  \\
\hline

\multirow{2}{*}{ $4$ } & $l\bar{l}$ &        $0$      &        $-$     &      $-$       &          $-$   &         $0$      &     $-$     &      $-$                    &      $-$   \\

                       & $s\bar{s}$ & $0.003\pm0.001$ &        $0$     &       $0$      &         $-1/3$ & $0.003\pm0.001$   &     $0$    & $0$                    & $-1/3$ \\
\hline

\multirow{2}{*}{ $5$ } & $l\bar{l}$ & $0$             &        $-$     &      $-$       &         $-$    &          $0$       &    $-$     &      $-$                    &         $-$    \\

                       & $s\bar{s}$ & $0.009\pm0.001$ &        $0$     &       $0$      &         $-1/3$ &  $0.009\pm0.001$   &     $0$    & $0$                    & $-1/3$\\
\hline

\multirow{2}{*}{ $6$ } & $l\bar{l}$ & $0.041\pm0.004$ &       $-2/9$   &     $-1/9$     &        $0$     &   $0.045\pm0.005$  &    $-2/9$  & $-1/9$                 & $0$ \\

                       & $s\bar{s}$ & $0$             &        $-$     &      $-$       &         $-$    &            $0$     &     $-$    &      $-$                    &         $-$    \\
\hline

\multirow{2}{*}{ $7$ } & $l\bar{l}$ & $0$             &        $-$     &      $-$       &         $-$    &            $0$     &     $-$    &      $-$                    &         $-$    \\

                       & $s\bar{s}$ & $0.007\pm0.001$ &        $0$     &       $0$      &         $-1/3$  & $0.007\pm0.001$    &     $0$   & $0$                    & $-1/3$\\
\hline


\multirow{2}{*}{ $8$ } & $l\bar{l}$ & $0.073\pm0.007$ &        $2/3$    &     $1/3$      &        $0$     & $0.157\pm0.016$     &  $4/9$   & $-1/9$                 & $0$\\

                       & $s\bar{s}$ & $0$             &         $-$     &      $-$       &         $-$    &        $0$          &   $-$    &      $-$                    &         $-$    \\
\hline

\multirow{2}{*}{ $9$ } & $l\bar{l}$ & $0$             &        $-$      &      $-$       &         $-$    &         $0$         &   $-$    &      $-$                    &         $-$    \\

                       & $s\bar{s}$ & $0.006\pm0.001$ &         $0$     &      $0$       &         $1$    &    $0.014\pm0.002$  & $4/9$    & $-1/9$                 & $0$ \\
\hline

\multirow{2}{*}{ $10$ }& $l\bar{l}$ & $0$             &        $-$     &      $-$       &         $-$    &        $0$          &   $-$    &      $-$                    &         $-$    \\

                       & $s\bar{s}$ & $0.003\pm0.001$ &         $0$     &      $0$       &         $1$   &  $0.007\pm0.001$    &   $1/2$  & $1/12$                 & $-1/4$\\
\hline

\multirow{2}{*}{ $11$ }& $l\bar{l}$ & $0.006\pm0.001$ &         $0$     &      $1$       &         $0$    & $0.013\pm0.001$    &  $1/3$ & $0$                    & $0$\\

                       & $s\bar{s}$ & $0.002\pm0.000$ &         $0$     &      $0$       &         $1$    &  $0.004\pm0.001$   &  $1/3$   & $1/6$                  & $-1/6$\\
\hline

\multirow{2}{*}{ $12$ }& $l\bar{l}$ & $0.005\pm0.001$ &         $2/3$   &      $1/3$     &         $0$    &    $0.010\pm0.001$ &  $1/3$   & $0$                    & $0$ \\

                       & $s\bar{s}$ & $0$             &         $-$     &      $-$       &         $-$    &         $0$         &  $-$    & $-$                    & $-$\\
\hline

\multirow{2}{*}{ $13$ }& $l\bar{l}$ & $0$             &        $-$     &      $-$       &         $-$    &         $0$         &  $-$    & $-$                    & $-$\\

                       & $s\bar{s}$ & $0.001\pm0.000$ &          $0$     &      $0$       &         $1$   &   $0.002\pm0.001$  &  $7/18$  & $1/36$                 & $-1/12$\\
\hline

\multirow{2}{*}{ $14$ }& $l\bar{l}$ & $0$             &         $-$     &      $-$       &         $-$    &         $0$         &  $-$    & $-$                    & $-$\\

                       & $s\bar{s}$ & $0.008\pm0.001$ &          $0$     &      $0$       &         $1$   &    $0.017\pm0.002$  &  $1/2$  & $1/12$                 & $-1/4$\\
\hline

\multirow{2}{*}{ $15$ }& $l\bar{l}$ & $0.015\pm0.002$ &           $0$    &      $1$       &         $0$   &    $0.032\pm0.003$  &  $1/3$  & $0$                    & $0$\\

                       & $s\bar{s}$ & $0.004\pm0.001$ &           $0$     &      $0$       &         $1$  &     $0.009\pm0.001$  &  $1/3$ & $1/6$                  & $-1/6$\\
\hline

\multirow{2}{*}{ $16$ }& $l\bar{l}$ & $0.012\pm0.001$ &         $2/3$     &      $1/3$     &         $0$   &     $0.025\pm0.002$  &  $1/3$ & $0$                    & $0$ \\

                       & $s\bar{s}$ & $0$             &         $-$     &      $-$       &         $-$    &         $0$         &  $-$    & $-$                    & $-$\\
\hline

\multirow{2}{*}{ $17$ }& $l\bar{l}$ & $0$             &         $-$     &      $-$       &         $-$    &         $0$         &  $-$    & $-$                    & $-$\\

                       & $s\bar{s}$ & $0.002\pm0.000$ &           $0$     &      $0$       &         $1$  &     $0.004\pm0.001$  &  $7/18$ & $1/36$                 & $-1/12$\\

\bottomrule[1.2pt]

\end{tabular}

\end{table*}

It's shown that $\Delta f_{i}^{q}$ are the same ones for the
five-quark configurations with spin symmetry $[22]^{S}$ in Set I and
Set II, this is because that both the two sets share the same
wave function of Eq.~(\ref{22s}) for these configurations. For
configurations numbered $i=8-17$, whose wave functions are
Eqs.~(\ref{31s0}) and~(\ref{31s1}) for Set I and II, respectively,
the obtained $\Delta f_{i}^{q}$ are therefore different. While the
probabilities for any configurations in the wave function of proton
should be different for the two sets. Indeed, for the case of $J=0$,
the total obtained probabilities for the five-quark components with
$u\bar{u}$, $d\bar{d}$ and $s\bar{s}$ pairs are
\begin{eqnarray}
P_{u\bar{u}} &=& 0.098\pm0.010 \,, \\
P_{d\bar{d}} &=& 0.216\pm0.022 \,,  \\
P_{s\bar{s}} &=& 0.057\pm0.006\, .
\end{eqnarray}
These numerical values yield the ratios
\begin{eqnarray}
r_{l} &=& \frac{P_{u\bar{u}}}{P_{d\bar{d}}} = 0.5 \pm 0.1 \,, \\
r_{s} &=& \frac{P_{s\bar{s}}}{P_{d\bar{d}}} = 0.3 \pm 0.1 \,, \\
\kappa_{s} &=& \frac{2P_{s\bar{s}}}{P_{u\bar{u}} + P_{d\bar{d}}}=
0.4 \pm 0.1 \,,
\end{eqnarray}
which can be related to the ratios of pseudoscalar meson
electro-production. These ratios were extracted to be $r_{l}=0.74
\pm 0.18$, $r_{s} = 0.22 \pm 0.07$ and $0.25 \pm 0.09$ by CLAS
collaboration~\cite{Park:2014zra}. It is found that, within errors,
our results are roughly consistent with these above experimental
data.

While for the case of $J=1$, the theoretical results are
\begin{eqnarray}
P_{u\bar{u}} &=& 0.170\pm0.017\, ,\\
P_{d\bar{d}} &=& 0.288\pm0.029\, ,\\
P_{s\bar{s}} &=& 0.090\pm0.009\,.
\end{eqnarray}
The probabilities for both light and strange five-quark Fock
components are larger than those obtained by Set I, which is because
of that the coupling strength between the five-quark configurations
with the four-quark spin symmetry $[31]^{S}$ and the traditional
three-quark component of proton for Set II is larger than that for
Set I. Accordingly, one can get the ratios
\begin{eqnarray}
r_{l} &=& 0.6 \pm 0.1\, ,\\
r_{s} &=& 0.3 \pm 0.1\, ,\\
\kappa_{s} &=& 0.4 \pm 0.1\,,
\end{eqnarray}
these values are similar with those obtained by Set I, but one has
to note that the total probability of the five-quark components in
Set II is relatively large, and the numerical results are in
agreement with both the $\bar{d}-\bar{u}$ and $\bar{u}/\bar{d}$ data
quite well. In addition, the present obtained probabilities of the
five-quark components with light quark-antiquark pair are very close
to the results obtained by the BHPS model that~\cite{Chang:2011vx}
\begin{equation}
P_{u\bar{u}}=0.176,~~~P_{d\bar{d}}=0.294\,.
\end{equation}

Next, we discuss in detail the theoretical results in the following
two subsections for Set I and Set II, respectively.

\subsection{Set I}
\label{set1}

From the quark spin matrix elements $\Delta f_{i}^{q}$ obtained from
the wave function as in Eqs.~(\ref{22s}) and~(\ref{31s0}) for Set I,
one can easily find that the total spin and total angular momentum
of the four-quark subsystem in the five-quark configurations with
four-quark orbital symmetry $[31]^\chi$ are $0$. For the
configurations with four-quark spin symmetry $[22]^{S}$ and orbital
symmetry $[4]^\chi$, and for those with four-quark spin symmetry
$[31]^{S}$ and orbital symmetry $[4]^\chi$, the total orbital
angular momentum of the quarks and the antiquark are also $0$. Thus,
for all the 17 five-quark configurations, only the antiquark spin
contributes to $\Delta f_{i}^{q}$.

With the obtained probabilities $P_{q\bar{q}}$ and the matrix
elements $\Delta f_{i}^{q}$, we get the total quark spin $\Delta f$
shown in Table~\ref{Delta}, and the axial charges of proton shown in
Table~\ref{nac}, compared to the experimental measurement by
COMPASS~\cite{Alekseev:2010hc} in the rows denoted by EXP, and the
latest predictions by lattice
QCD~\cite{Alekseev:2010hc,QCDSF:2011aa,Green:2017keo,Yamanaka:2018uud,Alexandrou:2020sml}
in the rows denoted by LQCD, predictions by the extended cloudy bag
model~\cite{Bass:2009ed} in the row denoted by CMB, and predictions
by the chiral perturbation theory~\cite{Li:2015exr} in the row
denoted by $\chi$PT.


\begin{table*}[htbp]
\caption{\footnotesize Numerical results of $\Delta f~(f=u,d,s)$,
compared to the data and predictions by lattice QCD and chiral
perturbation theory.} \label{Delta}
\renewcommand
\tabcolsep{0.45cm}
\renewcommand{\arraystretch}{1.8}
\scriptsize \vspace{0.5cm}
\begin{tabular}{ccccc}
\toprule[1.2pt]
                              &    $\Delta u$        &     $\Delta d$          &   $\Delta s$    \\
\hline

Set I                         &  $0.883\pm0.005$     &   $-0.213\pm0.003$     &  $0.015\pm0.002$ \\

Set II                        &  $0.710\pm0.012$    &   $-0.225\pm0.008$     &  $-0.020\pm0.003$ \\
\hline

EXP~\cite{Alekseev:2010hc}    & $0.84\pm0.01\pm0.02$ & $-0.430\pm0.01\pm0.02$  & $-0.030\pm0.01\pm0.02$\\

LQCD~\cite{QCDSF:2011aa}      & $0.794(21)(2)$       & $-0.289(16)(1)$         &    $-0.023(10)(1)$  \\

LQCD~\cite{Green:2017keo}     &   $0.863(7)(14)$     &  $-0.345(6)(9)$         & $-0.0240(21)(11)$\\

LQCD~\cite{Yamanaka:2018uud}  &                      &                         & $-0.046(26)(9)$\\

LQCD~\cite{Alexandrou:2020sml}&  $0.864(16)$         & $-0.426(16)$            & $-0.046(8)$ \\

$\chi$PT~\cite{Li:2015exr}    &  $0.90^{+0.03}_{-0.04}$    &  $-0.38^{+0.03}_{0.03}$ & $-0.007^{+0.004}_{-0.007}$\\

    \bottomrule[1.2pt]
\end{tabular}
\end{table*}


\begin{table*}[htbp]
\caption{\footnotesize Numerical results of $g^{(3)}_A,\,g^{(8)}_A
\text{ and }g^{(0)}_A$, compared to the data and predictions by
extended cloudy bag model, and lattice QCD and chiral perturbation
theory.} \label{nac}
\renewcommand
\tabcolsep{0.45cm}
\renewcommand{\arraystretch}{1.8}
\scriptsize \vspace{0.5cm}
\begin{tabular}{ccccc}
\toprule[1.2pt]
                            &  $g^{(3)}_A$            &          $g^{(8)}_A$          &  $g^{(0)}_A$        \\
\hline
Set I                       &   $1.096\pm0.0053$      &       $0.640\pm0.010$         &  $0.685\pm0.0076$  \\

Set II                      &   $0.935\pm0.0192$      &       $0.525\pm0.0258$        &    $0.465\pm0.0225$ \\
\hline

EXP~\cite{Alekseev:2010hc}  &   $1.2670(35)$          &      $0.58\pm0.03\pm0.05$     &     $0.33\pm0.03$  \\

LQCD~\cite{QCDSF:2011aa}    &     $1.082(18)(2)$      &        $0.550(24)(1)$         &     $0.482(38)(2)$\\

LQCD~\cite{Green:2017keo}   &   $1.208(6)(16)(1)(10)$ &       $0.565(11)(13)$         &  $0.494(11)(15)$ \\

LQCD~\cite{Yamanaka:2018uud}&  $1.123(28)(29)(90)$    &                               &            \\

CBM~\cite{Bass:2009ed}      &    $1.270$              &       $0.420\pm0.02$          &     $0.370\pm0.02$\\

$\chi$PT~\cite{Li:2015exr}  &     $1.27$              &       $0.53^{+0.06}_{-0.06}$  & $0.51^{+0.07}_{-0.08}$\\

    \bottomrule[1.2pt]
\end{tabular}
\end{table*}

As one can see in Table~\ref{Delta}, predictions for $\Delta u$ by
most of the theoretical approaches are about $0.8 \sim 0.9$ which
fall very well in the range of the COMPASS data, while those for
$\Delta d$ and $\Delta s$ show a little deviation from the data.
Similarly, the present obtained $\Delta u$ is very close to the
data, but $\Delta d$ is only about half of the experimental value,
while the worst result is that the sign of the present $\Delta s$ is
contrary to that of the experimental result, as well the predictions
by other approaches. As discussed above, in present case, only the
antiquark spin contributes to $\Delta f_{i}^{q}$, and the spin
states of the anti-strange quark in the configurations numbered
$8-17$ should be $|1/2,\uparrow\rangle$, therefore, we obtain a
small but positive value for $\Delta s$. Most of the predictions for
$\Delta s$ by other theoretical approaches are roughly consistent
with the COMPASS measurement, while a recent investigation employing
chiral perturbation theory showed a very close to $0$ but negative
value~\cite{Li:2015exr,Wang:2020hkn}.

In Table~\ref{nac}, the obtained $g_{A}^{(3)}$ and $g_{A}^{(8)}$ are
in agreement with the data and these predictions from other
approaches, with a $\sim 10\%$ deviation. But the present
$g_{A}^{(0)}$, which should indicate the quark spin contributions to
proton spin, is more than twice of the COMPASS data, this is because
of the small absolute value of $\Delta d$ and the positive $\Delta
s$ obtained in present model. In Ref.~\cite{Bass:2009ed}, a cloudy
bag model was applied to the axial charges of proton, the $N \pi$,
$\Delta \pi$ and $\Lambda K$ Fock components were taken into
account, by considering the relativistic corrections, they obtained
the numerical results for proton axial charges which fitted the
experimental data very well.

\subsection{Set II}
\label{set2}

For the quark spin matrix elements $\Delta f_{i}^{q}$,
differing from Set I, both the quarks
and antiquarks in the configurations numbered $8-17$ could contribute.
And with the numerical results for $P_{q\bar{q}}$ and $\Delta f_{i}^{q}$,
we get the results of the quark spin $\Delta f$ and the axial charges
of proton listed in Table~\ref{Delta} and~\ref{nac}, respectively, both results
are compared to the experimental data and predictions by other theoretical
approaches.

One can immediately find that the present obtained $\Delta u$ has a
deviation from the experimental data by more than $10\%$, and the
numerical value for $\Delta d$ is very close to the one obtained by
Set I, which is smaller than the COMPASS data. The $\Delta s$ is
small and negative. All these numerical results for $\Delta f$ are
close to the results published by QCDSF
collaboration~\cite{QCDSF:2011aa}.

For the isovector, flavor octet and singlet axial charges of the
proton, as one can see in Table~\ref{nac}, the obtained
$g_{A}^{(3)}$ is only about $3/4$ of the COMPASS data. This is
because a relatively small value of $\Delta u$ is obtained in
present Set II. While numerical values of the flavor octet axial
charge $g_{A}^{(8)}$ and the flavor singlet axial charge
$g_{A}^{(0)}$ are very close to the lattice QCD predictions in
Refs.~\cite{QCDSF:2011aa,Green:2017keo}.


\section{Summary}
\label{conc}

To summarize, in present manuscript, the intrinsic sea content in
proton is investigated using an extended chiral constituent quark
model in which all the five-quark Fock components are taken into
account. We take the $^{3}P_{0}$ quark-antiquark creation mechanism
to get the transition coupling strength between the three- and
five-quark components. Taking the empirical values for the model
parameters, one can get the total probability of all possible
five-quark components in proton wave function which is about $30
\sim 50\%$. The theoretical calculations can fit both the sea flavor
asymmetry $\bar{d}-\bar{u}$~\cite{Towell:2001nh} and the strangeness
suppression of $q\bar{q}$ creation obtained recently by CLAS
collaboration~\cite{Park:2014zra}.

In present phenomenological work, to form the positive parity of
proton, one of the quarks or antiquark in a given five-quark Fock
component of the proton wave function must be in the orbital
$P$-state, which contribute a value $\sim0.158$ to the spin of
proton~\cite{An:2019tld}. Consequently, the obtained wave function
for proton in present work is then applied to study the inner quark
spin of proton. Our numerical results show that the obtained $\Delta
u$, $\Delta d$ and $\Delta s$ are in consistent with the
experimental data and predictions by other theoretical approaches.

Within the obtained proton wave functions, we study the isovector,
flavor octet and singlet axial charges. It is shown that the
probabilities $P_{q\bar{q}}$ might reach to $\sim 50\%$ for the
five-quark Fock components in proton, and the resulted values for
$g_{A}^{(3)}$, $g_{A}^{(0)}$ and $g_{A}^{(8)}$ are consistent with
the predictions by other theoretical approaches. Note that we have
considered only the contributions from the inner quark spin, thus
the present obtained numerical results are not fully in agreement
with the experimental data. We will also study contributions from
the gluon spin and orbital angular momentum, in future, when more
experimental data are available.


\begin{acknowledgments}

This work is partly supported by the National Natural Science
Foundation of China under Grant Nos. 11675131, 12075288, 12075133, 11735003,
11961141004 and 11835015. It is also supported by the Youth Innovation
Promotion Association CAS, Taishan Scholar Project of Shandong Province (Grant No.tsqn202103062) and the Higher Educational Youth Innovation Science and Technology Program Shandong Province (Grant No. 2020KJJ004).

\end{acknowledgments}




\end{document}